\RequirePackage{ifpdf}
\ifpdf 
\documentclass[pdftex]{sigma}
\else
\documentclass{sigma}
\fi

\begin{document}
\allowdisplaybreaks

\renewcommand{\PaperNumber}{013}

\FirstPageHeading

\ShortArticleName{Operator Gauge Symmetry in QED}

\ArticleName{Operator Gauge Symmetry in QED}

\Author{Siamak KHADEMI~$^\dag$ and Sadollah NASIRI~$^\dag$$^\ddag$} 

\AuthorNameForHeading{S. Khademi and S. Nasiri}

\Address{$^\dag$~Department of Physics, Zanjan University, P.O.
Box 313, Zanjan, Iran}
\EmailD{\href{mailto:skhademi@mail.znu.ac.ir}{skhademi@mail.znu.ac.ir}}

\URLaddressD{\url{http://www.znu.ac.ir/Members/khademi.htm}}

\Address{$^\ddag$~Institute for Advanced Studies in Basic
Sciences, IASBS, Zanjan, Iran}
\EmailD{\href{mailto:nasiri@iasbs.ac.ir}{nasiri@iasbs.ac.ir}}

\ArticleDates{Received October 09, 2005, in f\/inal form January 17, 2006; Published online January 30, 2006}

\Abstract{In this paper, operator gauge transformation, f\/irst
introduced by Kobe, is applied to Maxwell's equations and
continuity equation in QED. The gauge invariance is satisf\/ied
after quantization of electromagnetic f\/ields. Inherent
nonlinearity in Maxwell's equations is obtained as a direct result
due to the nonlinearity of the operator gauge transformations. The
operator gauge invariant Maxwell's equations and corresponding
charge conservation are obtained by def\/ining the {\it generalized}
derivatives of the {\it f\/irst} and {\it second} kinds.
Conservation laws for the {\it real} and {\it virtual} charges are
obtained too. The additional terms in the f\/ield strength tensor
are interpreted as electric and magnetic polarization of the
vacuum.}


\Classification{81V80; 78A25}

\section{Introduction}
Conserved quantities of a system are direct consequences of
symmetries inherent in the system. Therefore, symmetries are
fundamental properties of any dynamical system \cite{1,2,3}. For 
a~system of electric charges an important quantity that is always
expected to be conserved is the total charge. The symmetry
operation under which the charge is conserved is known as gauge
transformation \cite{4,5,7,8}.

In classical and semi-classical electrodynamics all gauges are
equivalent, and one commonly pre\-supposes a gauge condition on
the electromagnetic (EM) potentials to simplify
calcula\-tions~\mbox{\cite{9,10}}. The gauge invariance in QED is
extensively studied by Kobe \cite{11,12}. The behavior of
propagators for quantum electrodynamics and their equivalence in
dif\/ferent gauges is consi\-dered by Zumino \cite{13}, Gaete
\cite{14}, Manoukian \cite{15} and Manukian and Siranan \cite{16}.
In this paper, the charge conservation is shown to be recovered
after quantizing the EM f\/ields as well. This requires that the EM
f\/ields and Maxwell's equations to be gauge invariant in quantum
level. The conventional approach to quantum electrodynamics (QED)
assumes the gauge f\/ixing before the quantization procedure
\cite{17}. In this procedure the EM potentials are promoted to
operators and gauge conditions are expressed as operator
relations. Thus, in QED, instead of simple gauge transformations,
one has to deal with more complicated operator gauge
transformations (OGTs)~\cite{18}. However, it is shown that the
ordinary EM f\/ields and Maxwell's equations become gauge dependent
in this method \cite{19}. This inconsistency is extensively
studied by dif\/ferent authors \cite{20,21,22}. To overcome this
problem, Kobe \cite{18} quantized the EM f\/ields without gauge
f\/ixing and redef\/ined the EM f\/ields strength tensor that is
invariant under the OGTs. However, it can be easily shown that
having the operator gauge invariance of EM f\/ields strength tensor
is not suf\/f\/icient for Maxwell's equations to be operator gauge
invariant.

Here, we extend the Kobe approach to obtain the operator gauge
invariant Maxwell's equations and corresponding charge
conservation. It seems that the discrepancy of the gauge symmetry
mentioned before is removed and  the gauge invariance is preserved
under the dif\/ferent OGTs. The non-linear gauge transformation as a
subset of non-linear transformations is investigated by Doebner et
al \cite{23} and Goldin et al \cite{24}. They describe an
inherently linear system by non-linear dynamical equations. Thus,
the non-linearity in Maxwell's equations obtained in this paper is
a~direct consequence of the fact that the operator gauge
transformations are non-linear transformations. Furthermore, the
classical electromagnetic f\/ields as Abelian gauge f\/ields are
promoted to non-Abelian gauges in QED. Thus, the operator gauge
invariant Maxwell's equations become mathematically similar to
Yang--Mills f\/ields. The additional commutators in quantized
electric and magnetic f\/ields are expressed as electric and
magnetic polarization, although they are gauge dependent.

The layout of this paper is as follows. In Section~2, after a
brief review of semi-classical electrodynamics as well as the
corresponding gauge transformations, the concept of operator gauge
transformations is introduced. The generalized Maxwell's equations
and conservation law for electric charge are given. In Section~3,
the polarization of vacuum is discussed and the last section  is
devoted to conclusions and results.

\section{Operator gauge invariant formulation of QED}

In semi-classical electrodynamics the charged particles are
quantized and their dynamics is described by the Schr\"odinger
equation
\begin{gather}
H|\psi({\boldsymbol r},t)\rangle=i\hbar\frac{\partial}{\partial
t}|\psi({\boldsymbol r},t)\rangle,\label{eq1}
\end{gather}
while the EM f\/ields are still classical. In equation \eqref{eq1}
one has
\begin{gather*}
H=\frac{1}{2m}\left({\boldsymbol P}-\frac{q}{c}{\boldsymbol
A}\right)^{2}+q\phi +V,
\end{gather*}
where $A_{\mu}=(\phi ,-{\boldsymbol A})$ and $V$ is
non-electromagnetic potentials. For charged particles, the
expectation value of any physical quantity $G$ may be obtained by
the averaging of the corresponding quantum mechanical operator on
the state ket $|\psi({\boldsymbol r},t)\rangle$ as
\begin{gather*}
\langle G\rangle =\langle \psi|G|\psi\rangle.
\end{gather*}

The gauge transformations are described by unitary transformations
as follows
\begin{gather}
S=\exp\left(\frac{iq\Lambda({\boldsymbol r},t)}{\hbar
c}\right),\label{eq2}
\end{gather}
which acts on the state kets as
\begin{gather}
|\psi^{\prime}\rangle=S|\psi \rangle.\label{eq3}
\end{gather}

In equation \eqref{eq2} $\Lambda=\Lambda({\boldsymbol r},t)$ is an
arbitrary smooth gauge function. If the form invariance of the
Schr\"odinger equation under such unitary transformations is
asserted, then the components of corresponding gauge transformed
four potential will be
\begin{gather}
A^{\prime}_{\mu}=A_{\mu}+\partial_{\mu} \Lambda.\label{eq4}
\end{gather}
In QED, in addition to the charged particles, potentials of EM
f\/ields (thereby, the EM f\/ields themselves) are also quantized.
That is, such entities must be promoted to operators, as well. In
this case Maxwell's equations are also promoted to operator
equations.

In general, the gauge function in equation \eqref{eq2} is a
function of  EM potentials $A_{\mu}$ and para\-me\-ters~$x^{\nu}$
\begin{gather*}
\Lambda=\Lambda\big(x^{\nu},A_{\mu}(x^{\nu})\big).
\end{gather*}
Therefore, $\Lambda$ is promoted to a Hermitian operator, because
$A_{\mu}$ is now an operator. Equation~\eqref{eq3} now becomes
\begin{gather}
|\psi^{\prime}\rangle=\exp\left(\frac{iq\Lambda(x^{\nu},A_{\mu}(x^{\nu}))}{\hbar
c}\right)|\psi\rangle.\label{eq5}
\end{gather}
When the form invariance of the Schr\"odinger equation is assumed,
one may f\/ind
\begin{gather}
H^{\prime}=H(A^{\prime}_{\mu})=SHS^{-1}+i\hbar
S\left(\frac{\partial S^{-1}}{\partial t}\right),\label{eq6}
\end{gather}
where the Hamiltonian of a charged particle is given by
equation~\eqref{eq1}. The operator gauge transformed potentials in
equation~\eqref{eq6} can be shown to be
\begin{gather}
A^{\prime}_{\mu}=SA_{\mu}S^{-1}+\left(\frac{i\hbar
c}{q}\right)S\big(\partial_{\mu}S^{-1}\big).\label{eq7}
\end{gather}
Note that the OGT def\/ined by equation~\eqref{eq7} is quite
dif\/ferent from the ordinary gauge transformations, given by
\eqref{eq4}. This is due to the fact that, in general,
\begin{gather*}
[A_{\mu},\Lambda (A_{\nu}(x^{\rho}))]\neq 0
\end{gather*}
and
\begin{gather*}
[\Lambda(A_{\mu}),\dot{\Lambda}(A_{\nu})]\neq 0.
\end{gather*}
The generalized gauge transformation of equation~\eqref{eq7}
reduces to that of equation \eqref{eq4} in the classical and
semi-classical limits.

Since the gauge symmetry is a fundamental concept in QED, all
physical quantities and dynamical equations of particles and EM
f\/ields must be gauge invariant. The promotion of QED after a gauge
f\/ixing violates the gauge symmetry. However, introducing the OGTs
into QED, f\/irst done by Kobe~\cite{18}, ensures the gauge symmetry
in quantum level.

Further, a physical quantity, by its very meaning, must be unique.
Therefore, for an arbitrary operator, $G$, to represent an
observable, we demand that
\begin{gather}
\langle \psi^{\prime}|G^{\prime}|\psi^{\prime}\rangle =\langle
\psi|G|\psi\rangle.\label{eq8}
\end{gather}
Using equations \eqref{eq5} and \eqref{eq8} one may obtain the OGT
of any observable operator $G$ as follows
\begin{gather}
G^{\prime}(A_{\mu})=G(A^{\prime}_{\mu})=SG(A_{\mu})S^{-1}.\label{eq9}
\end{gather}
As an example it is clear that the Hamiltonian is operator gauge
dependent and cannot represent the observable energy, except when
$\frac{\partial S^{-1}}{\partial t}=0$. However, it is well known
that one may def\/ine the energy operator as ${\cal E}=H-q\phi$,
which transforms as equation~\eqref{eq9}, under OGT. In the same
manner, the f\/ield strength tensor $F_{\mu
\nu}=\partial_{\mu}A_{\nu}-\partial_{\nu}A_{\mu}$ now becomes an
operator which does not satisfy equation~\eqref{eq9} under OGT,
i.e.
\begin{gather*}
F^{\prime}_{\mu \nu}\neq SF_{\mu \nu}S^{-1}.
\end{gather*}
Consequently, $F_{\mu\nu}$ can not be a physical quantity in the
framework of the operator gauge invariant QED. To remedy this
inconsistency, one may redef\/ine the EM f\/ield strength tensor as
\begin{gather}
F_{\mu\nu}=\partial_{\mu}A_{\nu}-\partial_{\nu}A_{\mu}+\frac{iq}{\hbar
c}[A_{\mu},A_{\nu}],\label{eq10}
\end{gather}
which may be easily shown to be operator gauge invariant
\cite{18}.

The classical equations of EM f\/ields, i.e., the Maxwell equations,
should be transformed into operators, when the EM f\/ields are
quantized. Clearly, these equations are operator gauge dependent,
therefore, it violates the gauge symmetry principle, in QED. To
overcome this problem we replace the four-derivative operator
$\partial \mu$ in ordinary def\/inition of $F_{\mu\nu}$ by {\it
generalized} four-derivative as follows
\begin{gather}
\partial_{\mu}\rightarrow D_{\mu}=\partial _{\mu}+\frac{iq}{\hbar c}A_{\mu},\label{eq11}
\end{gather}
to ensure the gauge invariance of physical quantities in QED. From
equation~\eqref{eq11} it can be shown that the gauge
transformation of the generalized four-derivative in QED obeys
\begin{gather*}
D^{\prime}_{\mu}=\partial_{\mu}+\frac{iq}{\hbar
c}A_{\mu}^{\prime}=S\left\{\partial_{\mu}+\frac{iq}{\hbar
c}A_{\mu}\right\} S^{-1}=SD_{\mu}S^{-1}.
\end{gather*}
Therefore, one must use the generalized four-derivative of the
four potentials instead of the common derivative to def\/ine
$F_{\mu\nu}$, in an operator gauge invariance formulation of QED
\begin{gather}
F_{\mu\nu}=D_{\mu}A_{\nu}-D_{\nu}A_{\mu}.\label{eq12}
\end{gather}
We call $D_\mu$ as the generalized four-derivative of {\it f\/irst
kind}.

Equation \eqref{eq12} guarantees the operator gauge invariance of
$F_{\mu\nu}$. However, this is not suf\/f\/icient to preserve the
gauge symmetry of the Maxwell equations in QED. To get
consistency, we introduce the generalized four-derivative of the
{\it second kind} as follows
\begin{gather*}
{\cal D}_{\mu}=\partial_{\mu}+\frac{iq}{\hbar c}[A_{\mu},\ \  ],
\end{gather*}
which operates {\it only} on the {\it observable} as
\begin{gather*}
{\cal D}_{\mu}G=\partial_{\mu}G+\frac{iq}{\hbar c}[A_{\mu},G].
\end{gather*}

Thus, Maxwell's equations regarded as dynamical equations
governing the evolution of the observable in QED, take the
operator gauge invariant form as follows
\begin{gather}
{\cal D}_{\mu}F^{\mu\nu}=\partial_{\mu}F^{\mu\nu}+\frac{iq}{\hbar
c}[A_{\mu},F^{\mu\nu}]=\frac{1}{c}j^{\nu}\label{eq13}
\end{gather}
and
\begin{gather*}
{\cal D}_{\mu}
{}^{*}F^{\mu\nu}=\partial_{\mu}{}^{*}F^{\mu\nu}+\frac{iq}{\hbar
c}[A_{\mu},{}^{*}F^{\mu\nu}]=0,
\end{gather*}
where $^{*}F^{\mu \nu}=\frac{1}{2}\epsilon^{\mu \nu \alpha
\beta}F_{\alpha \beta}$ is the dual tensor corresponding to
$F^{\mu\nu}$. For a special class of gauge operators, where the
above commutators vanish, one ends up with ordinary Maxwell's
equations. One may call this special class of gauge  operators as
{\it commutative gauges}, otherwise, we have {\it non-commutative
gauges}.

Furthermore, if one takes the four-divergence of ordinary
inhomogeneous Maxwell's equation, and notes that
$F_{\mu\nu}=-F_{\nu\mu}$, one f\/inds,
\begin{gather*}
\partial_{\mu}j^{\mu}=0,
\end{gather*}
which is the conservation law of the electric charge. However, in
an operator-gauge-invariant formulation of QED, taking the 
four-divergence of equation~\eqref{eq13}, one  f\/inds
\begin{gather}
\partial_{\nu}\partial_{\mu}F^{\mu\nu}=\frac{1}{c}\partial_{\nu}{\cal J}^{\nu}=0,\label{eq14}
\end{gather}
where
\begin{gather}
{\cal
J}^{\nu}=j^{\nu}-\frac{iq}{\hbar}[A_{\mu},F^{\mu\nu}]\label{eq15}
\end{gather}
is the total {\it unobservable} four-current density. The f\/irst
term on the right-hand side of equation~\eqref{eq15} has its
origin in charged particle which produces the {\it real} (gauge
independent) four-current density, while the second term has only
the characteristic of the EM f\/ields, which produces the {\it
virtual} (gauge dependent) four-current density. Equation
\eqref{eq14} conf\/irms the conservation law of both real and
virtual charges. If one considers that the real four-current
density is an observable quantity, then
\begin{gather}
{\cal D}_{\mu}j^{\mu}=\partial_{\mu}j^{\mu}+\frac{iq}{\hbar
c}[A_{\mu},j^{\mu}]=0\label{eq16}
\end{gather}
will be the operator gauge invariant conservation law for real
charge. Note that ${\cal D}_{\mu}$ is the second kind derivative
which operates on observable four current density. From
equations~\eqref{eq14}--\eqref{eq16} one f\/inds
\begin{gather}
[A_{\mu},j^{\mu}]=\partial_{\nu}[A_{\mu},F^{\mu\nu}].\label{eq17}
\end{gather}
Therefore, whenever the commutator of $A_{\mu}$ and $j^{\mu}$
vanishes, equation~\eqref{eq17} gives the conservation of virtual
charge
\begin{gather}
\partial_{\nu}[A_{\mu},F^{\mu\nu}]=0.\label{eq18}
\end{gather}

Since the EM four-potentials and the four-current density act in
dif\/ferent ket spaces, the commutator on the left-hand side of
equation~\eqref{eq17} vanishes. Therefore, for a charged particle
interacting with an EM f\/ield, the virtual charge, as well as, the
real charge are conserved.

\section{Vacuum polarization}

Note that equation~\eqref{eq10}, has two operator gauge dependent
terms, while their combination is operator gauge independent. The
operator gauge independent electric and magnetic f\/ields, as the
elements of the real f\/ield strength tensor are as follows
\begin{gather}
(E_i)_{\rm new}=(E_i)_{\rm old}-\frac{iq}{\hbar
c}[\phi,A_i]\label{eq19}
\end{gather}
and
\begin{gather}
(B_k)_{\rm new}=(B_k)_{\rm old}-\frac{iq}{\hbar
c}\epsilon_{ijk}[A^i,A^j].\label{eq20}
\end{gather}
Consider the electric and magnetic polarization in a material
medium def\/ined by
\begin{gather}
\frac{1}{\epsilon_0}D_i=E_i+\frac{1}{\epsilon_0}P_i\label{eq21}
\end{gather}
and
\begin{gather*}
\mu_0 H_k=B_k-\mu_0 M_k.
\end{gather*}
In full QED one has a combination of
equations~\eqref{eq18}--\eqref{eq21} as
\begin{gather*}
\frac{1}{\epsilon_0}(D_i)_{\rm new}=(E_i)_{\rm old}-\frac{iq}{\hbar c}[\phi,A_i]
+\frac{1}{\epsilon_0}P_i=(E_i)_{\rm old}+\frac{1}{\epsilon_0}(P_i)_{\rm new}
\end{gather*}
and
\begin{gather*}
\mu_0 (H_k)_{\rm new}=(B_k)_{\rm old}-\frac{iq}{\hbar  c}\epsilon_{ijk}[A^i,A^j]-\mu_0 M_k
=(B_k)_{\rm old}-\mu_0 (M_k)_{\rm new},
\end{gather*}
 where
\begin{gather}
(P_i)_{\rm new}=(P_i)_{\rm old}-\frac{iq\epsilon_0}{\hbar
c}[\phi,A_i]=(P_i)_{\rm old}+(P_i)_{\rm vacuum} \label{eq22}
\end{gather}
and
\begin{gather}
(M_k)_{\rm new}=(M_k)_{\rm old}+\frac{iq}{\hbar
c\mu_0}\epsilon_{ijk}[A^i,A^j]=(M_k)_{\rm old}+(M_k)_{\rm
vacuum}.\label{eq23}
\end{gather}
In equations~\eqref{eq22} and \eqref{eq23} $(P_i)_{\rm vacuum}$
and $(M_k)_{\rm vacuum}$ denote the electric and magnetic
pola\-ri\-za\-tion of vacuum. Of course, the polarizations due to
vacuum are gauge dependent, therefore, are not measurable.
Whenever one chooses an special gauge in which the commutators in
equations~\eqref{eq19} and \eqref{eq20} vanish, then, the electric
and magnetic polarization become hidden, i.e., the old  and new
electric and magnetic f\/ields become identical.

\section{Conclusions}
Here we emphasize that the ordinary gauge symmetry of classical
electrodynamics to be preserved after quantization. Therefore, the
concept of gauge transformations as a basic symmetry of classical
electrodynamics, is extended to QED to obtain the  OGT law of EM
potentials. Expectation value of any observable, is then required
to be operator gauge invariant. In this respect, a more general
def\/inition for the observable is given. The ordinary EM f\/ield
strength tensor, which is an operator in QED, does not satisfy
this def\/inition and thus is no longer an observable. Therefore,
the operator gauge invariant form of this tensor is required to be
redef\/ined. Using this def\/inition, which is borrowed from Kobe,
shows that the ordinary Maxwell equations and charge conservation,
become operator gauge dependent. By def\/ining the generalized
four-derivatives of f\/irst and second kind, these dynamical
equations are consistently expressed in an operator gauge
invariant form, too. The conservation of the {\it real}, i.e.,
gauge independent, as well as the {\it virtual}, i.e., gauge
dependent charges are shown to be satisf\/ied. Derivation of vacuum
electric and magnetic polarizations as  direct consequence of
operator gauge symmetry of the formalism, emerges to be operator
gauge dependent quantities. Since the operator gauge
transformations are, in general, non-linear gauge transformations,
the operator gauge invariant Maxwell's equations become
non-linear, as well.   Furthermore, the Abelian gauge f\/ields of
classical Maxwell's equations are promoted to the non-Abelian
gauges due to the operator gauge symmetry.

\LastPageEnding
\end{document}